\documentclass{article}

     \usepackage[preprint,nonatbib]{neurips_2020}

\usepackage[utf8]{inputenc} 
\usepackage[T1]{fontenc}    
\usepackage{hyperref}       
\usepackage{url}            
\usepackage{booktabs}       
\usepackage{amsfonts}       
\usepackage{nicefrac}       
\usepackage{microtype}      
\usepackage{graphicx}       
\usepackage{subcaption}     

\title{Fibrosis-Net: A Tailored Deep Convolutional Neural Network Design for Prediction of Pulmonary Fibrosis Progression from Chest CT Images}

\author{Alexander Wong$^{1,2,3,*}$, Jack Lu$^{3}$, Adam Dorfman$^{3}$, Paul McInnis$^{3}$ \\
\textbf{Mahmoud Famouri$^{3}$, Daniel Manary$^{3}$,  James Ren Hou Lee$^{3}$, and Michael Lynch$^{3}$
} \\
			$^{1}$ Vision and Image Processing Research Group, University of Waterloo\\
			$^{2}$ Waterloo Artificial Intelligence Institute\\
			$^{3}$ DarwinAI Corp.\\ 			
			\texttt{$^{*}$a28wong@uwaterloo.ca}
	}

\begin{document}

\maketitle

\begin{abstract}
Pulmonary fibrosis is a devastating chronic lung disease that causes irreparable lung tissue scarring and damage, resulting in progressive loss in lung capacity and has no known cure. A critical step in the treatment and management of pulmonary fibrosis is the assessment of lung function decline, with computed tomography (CT) imaging being a particularly effective method for determining the extent of lung damage caused by pulmonary fibrosis.  Motivated by this, we introduce Fibrosis-Net, a deep convolutional neural network design tailored for the prediction of pulmonary fibrosis progression from chest CT images. More specifically, machine-driven design exploration was leveraged to determine a strong architectural design for CT lung analysis, upon which we build a customized network design tailored for predicting forced vital capacity (FVC) based on a patient's CT scan, initial spirometry measurement, and clinical metadata.  Finally, we leverage an explainability-driven performance validation strategy to study the decision-making behaviour of Fibrosis-Net as to verify that predictions are based on relevant visual indicators in CT images.  Experiments using a patient cohort from the OSIC Pulmonary Fibrosis Progression Challenge showed that the proposed Fibrosis-Net is able to achieve a significantly higher modified Laplace Log Likelihood score than the winning solutions on the challenge.  Furthermore, explainability-driven performance validation demonstrated that the proposed Fibrosis-Net exhibits correct decision-making behaviour by leveraging clinically-relevant visual indicators in CT images when making predictions on pulmonary fibrosis progress.  Fibrosis-Net is available to the general public in an open-source and open access manner\footnote{\url{https://github.com/darwinai/FibrosisNet}} as part of the OpenMedAI initiative. While Fibrosis-Net is not yet a production-ready clinical assessment solution, we hope that its release will encourage researchers, clinicians, and citizen data scientists alike to leverage and build upon it.
\end{abstract}
\section{Introduction}
Pulmonary fibrosis is a serious chronic lung disease in which permanent tissue scarring and damage occurs in the lungs.  The increasing replacement of healthy lung tissues with fibrotic tissue results in progressive, irreversible reduction in lung function over time.  There are currently no known cure and limited treatment options for pulmonary fibrosis, with treatment and management of the disease focused on the attenuation of lung function decline progression and improving quality of life.  The rate of progression for a patient with pulmonary fibrosis is highly variable, ranging from little to no change over many years to rapid deterioration in a short period of time.

A very critical step in the treatment and management of pulmonary fibrosis is the assessment of lung function decline, as this guides clinicians to determining the best course of treatment and management ranging from oxygen therapy and pulmonary rehabilitation to pharmacological agents (pirfenidone~\cite{Taniguchi} and nintedanib~\cite{Richeldi}) and lung transplantation~\cite{Kistler}. Guidelines set out by the ATS/ERS/JRS/ALAT \cite{raghu2018diagnosis} describe a number of methods for diagnosis of pulmonary fibrosis. Invasive techniques, such as surgical lung biopsy, have associated risks to the health and lung function for the patient \cite{richeldi2017idiopathic}. Transbroncial lung biopsy is a less invasive technique where small samples of the lung tissue are taken using video-assisted thoracoscopy or flexible bronchoscopy \cite{tomassetti2016bronchoscopic}. A number of methods have been utilized by clinicians for assessing lung function decline after diagnosis. For example, spirometry tests are frequently leveraged for measuring the FVC of the lung which is a key indicator of lung function ~\cite{Watters,Russell,Wuyts,du2011forced}.

One of the most effective methods for assessing lung function decline and the extent of lung damage due to pulmonary fibrosis is computed tomography (CT) imaging.  Several visual signs in CT scans have been identified and leveraged by radiologists to assess lung function decline from pulmonary fibrosis, with the most common visual indicator being honeycombing, which present as cystic spaces with irregularly thickened fibrotic tissue walls~\cite{Devaraj}.  However, given the rate of progression of different patients can be highly variable, the ability to accurately predict the progress of pulmonary fibrosis remains a major challenge.  This is further compounded by the fact that some common visual indicators such as honeycombing may not be present at certain stages of progression or even at all~\cite{Gruden}, while other atypical patterns mimicking other diseases may be present in the CT scans instead (e.g., predominance of ground-glass opacity, consolidation, nodules, and atypical distribution of lesions~\cite{Souza}, as well as the presence of ground-glass attenuation~\cite{Lynch}).   As such, new methods for improving prediction accuracy when leveraging CT images as a tool for assessing lung function decline due to pulmonary fibrosis is highly desired.

Motivated by the potential of machine learning for computer-aided clinical decision support for pulmonary fibrosis, in this study we introduce Fibrosis-Net, a deep convolutional neural network design tailored specifically for the prediction of pulmonary fibrosis progression from chest CT images. More specifically, machine-driven design exploration was leveraged to determine a strong architectural design for CT lung analysis, upon which we build a customized network design tailored for predicting forced vital capacity (FVC) based on a patient's CT scan, initial spirometry measurement, and clinical metadata.  Furthermore, to explore the decision-making behaviour of Fibrosis-Net, we leverage an explainability-driven performance validation strategy to audit Fibrosis-Net to verify that predictions are based on relevant visual indicators in CT images.  Fibrosis-Net is available to the general public in an open-source and open access manner\footnote{\url{https://github.com/darwinai/FibrosisNet}} as part of the OpenMedAI initiative, an open source initiative for medical artificial intelligence solutions that currently include the COVID-Net~\cite{covidnet,covidnets,covidnetct,covidnetct2,ebadi2021covidxus} initiative, Cancer-Net~\cite{lee2020cancernetsca} initiative, and the TB-Net initiative~\cite{wong2021tbnet}. While Fibrosis-Net is not yet a production-ready screening solution, we hope that its open source release will encourage researchers, clinicians, and citizen data scientists alike to leverage and build upon them.

The paper is organized as follows.  Related work in the area of artificial intelligence for computer-aided clinical decision support for pulmonary fibrosis is discussed in Section~\ref{relatedwork}.  Section~\ref{methods} provides a detailed description of the data preparation and analysis process, the architecture design construction process, the proposed Fibrosis-Net network architecture design, and the explainability-driven performance validation process.  Section~\ref{results} presents both the quantitative performance validation results evaluating the efficacy of the proposed Fibrosis-Net using a patient cohort from the OSIC Pulmonary Fibrosis Progression Challenge~\cite{OSIC} as well as visual validation results from the explainability-driven performance validation process used to study the decision-making behaviour of Fibrosis-Net.  Finally, Conclusions are drawn and future work is discuss in Section~\ref{conclusion} and the broader impact of the proposed work is discussed.

        \begin{figure}[t]
        \centering
        \includegraphics[width=\textwidth]{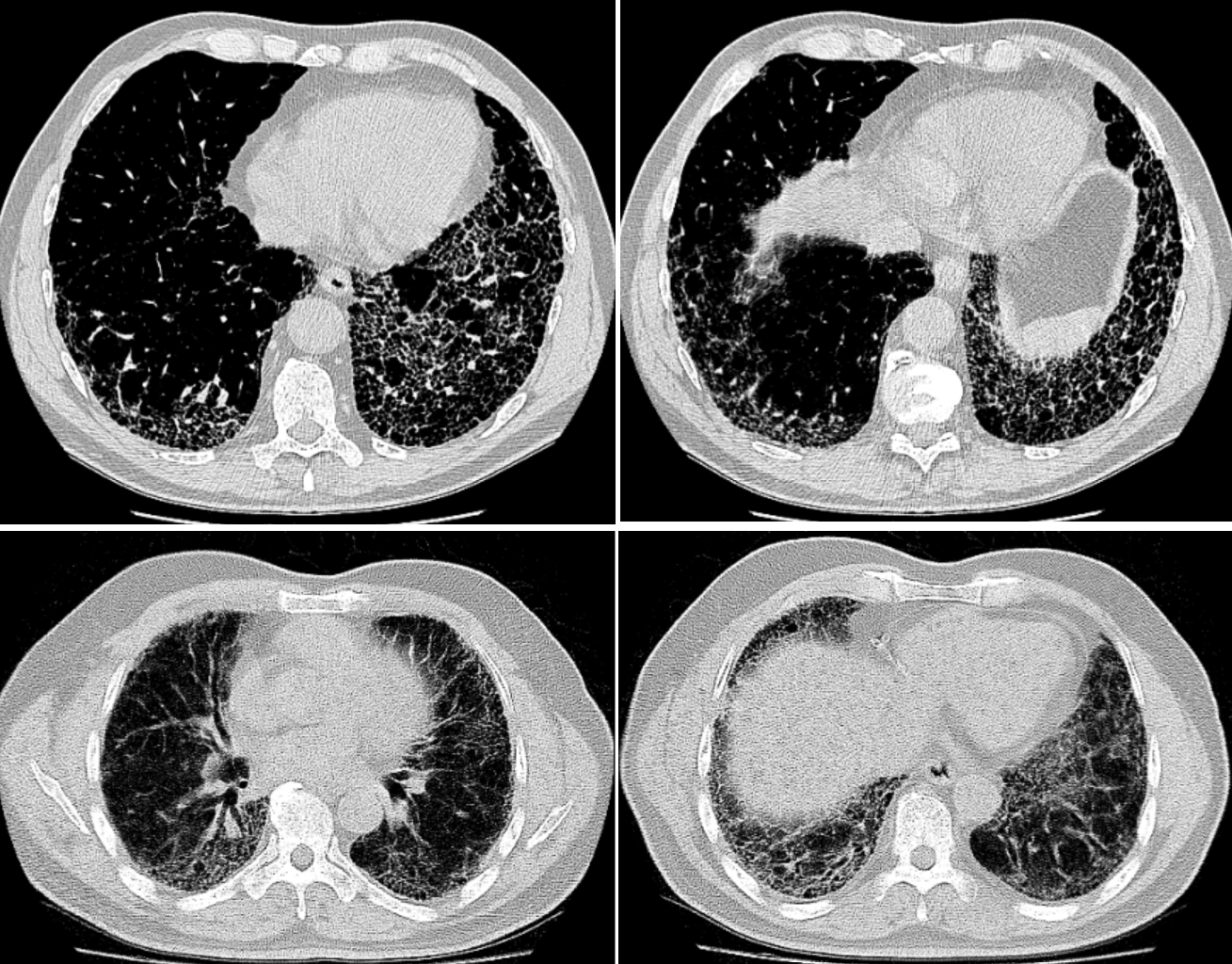}
        \caption{Example CT slices from the patient cohort from the OSIC Pulmonary Fibrosis Progression Challenge~\cite{OSIC}. }
        \label{fig:examples}
    \end{figure}

\section{Related work}\label{relatedwork}

Motivated by the significant benefits that can be gained and the challenges involved from a clinical perspective, there has been a recent interest in leveraging artificial intelligence for computer-aided clinical decision support of pulmonary fibrosis based on CT images~\cite{Christe,Walsh1,Walsh2,Levin}. For instance, Anthimopoulos et al. \cite{anthimopoulos2016lung} use a deep convolutional neural network to analyse 2D patches from the CT image to classify reticulation, honeycombing, ground glass opacity (GGO), consolidation, and micronodules in lung tissue. Christodoulidis et al. \cite{christodoulidis2016multisource} proposed a multi-source transfer learning approach with deep convolutional neural networks pre-trained with a selection of texture data sets, again with the goal of classifying 2D image patches of lung tissue in CT. Bermejo-Pel{\'a}ez et al. \cite{bermejo2020classification} describe a method using an ensemble of deep convolutional neural networks where the output of each network is summed up and weighted before being combined to form the overall output of the ensemble.

More recently, the significant potential and need for advancements in artificial intelligence-driven methods for computer-aided clinical decision support of pulmonary fibrosis was exemplified by the Kaggle Pulmonary Fibrosis Progression Challenge~\cite{OSIC} launched by the Open Source Imaging Consortium (OSIC) to get the research community to accelerate advancement of machine learning for pulmonary fibrosis assessment.  To the best of the authors' knowledge, the patient cohort curated by OSIC for the challenge is the largest publicly available cohort in literature.  Amongst the many artificial intelligence solutions introduced as part of the challenge, the 1st place winning solution~\cite{OSIC} proposed a weighted ensemble between a deep convolutional neural network with a state-of-the-art EfficientNet-B5 network architecture design~\cite{tan2020efficientnet} and a multiple quantile regressor to predict the lung function decline of a patient based on a patient's CT scans, initial spirometry measurement, and clinical metadata.

To the best of the authors' knowledge, the notion of explainability-driven performance validation of the decision-making behaviour of artificial intelligence solutions for the prediction of pulmonary fibrosis progression have not been previously explored in literature and can be very valuable for driving greater clinical adoption of such solutions in a transparent and trusted manner.  Furthermore, while explainability-driven performance validation strategies has been demonstrated to be very successful in past studies for the purpose of clinical classification~\cite{covidnet,covidnetct,covidnetct2,wong2021tbnet}, the the best of the authors' knowledge this is the first study in literature to successfully leverage explainability-driven performance validation on a clinical regression problem.

\section{Methods}\label{methods}
\subsection{Data preparation and analysis}
    To build the proposed Fibrosis-Net, we leverage the patient cohort from the OSIC Pulmonary Fibrosis Progression Challenge~\cite{OSIC}.  The data for this patient cohort consists of chest CT scans, forced vital capacity (FVC) measurements from frequent visits over the course of around 1-3 years, and associated clinical metadata (i.e., age, sex, smoking status, and patient's relative FVC measurement compared to the typical FVC measurement of a patient with similar characteristics).  More specifically, in this study, the training set consists of 172 patient cases, while the test set consists of approximately 28 patient cases.  To the best of the authors' knowledge, the patient cohort curated for this challenge is the largest publicly available cohort in literature.

Table~\ref{tab:demographic} summarizes the demographic variables of the data from  the patient cohort in the OSIC Pulmonary Fibrosis Progression Challenge~\cite{OSIC} used in this study. A number of observations can be made based on the demographic distribution analysis.  First, it can be observed that the patient cases in the cohort are distributed across the different age groups, with the mean age being 67.1 and the highest number of patients in the cohort are between the ages of 60-69.  This distributional trend towards older adults is reflective of the fact that pulmonary fibrosis is typically diagnosed later in life due to the condition worsening as time goes by.  Second, it can be observed that the majority of patient cases in the cohort are individuals who have smoked sometime in their life, with the majority of patients being ex-smokers.  This distributional trend is reflective of the fact that a recognized risk factor for the development of pulmonary fibrosis is smoking.  Furthermore, smoking can lead to more significant detrimental effects that can reduce the survival rate of patients with pulmonary fibrosis.  Third, it can be observed that a majority of the patient cases in the cohort are male.  This is consistent with clinical studies in literature showing that gender differences in the risk of pulmonary fibrosis~\cite{Ekstrom} as well as at clinical presentation~\cite{Kalafatis}.  Finally, it can be observed that the patient cases span a wide range of relative FVC measurement values, which indicates different levels of lung function decline.  This distribution diversity in relative FVC measurement values is desirable as it allows the deep neural network being trained to be exposed to a wider variety of pulmonary fibrosis progression scenarios for greater generalizability for different patient conditions.

\begin{table}[t]
\centering
\caption{Summary of demographic variables of data used in this study.}
\label{tab:demographic}
\begin{tabular}{|c|c|c|}
\hline
\multicolumn{1}{|l|}{\textbf{Age}} & mean $\pm$ std & $67.14 \pm 7.01$ \\ \hline
 & 50-59 & 14.9\% \\ \hline
 & 60-69 & 47.9\% \\ \hline
 & 70-79 & 34.4\% \\ \hline
 & 80-89 & 2.8\% \\ \hline
\multicolumn{3}{|l|}{\textbf{Sex}} \\ \hline
 & Male & 78.55\% \\ \hline
 & Female & 21.45\%\\ \hline
\multicolumn{3}{|l|}{\textbf{Smoking status}} \\ \hline
 & Currently smokes & 5.4\% \\ \hline
 & Ex-smoker & 66.3\% \\ \hline
 & Never smoked & 28.3\% \\ \hline
\multicolumn{3}{|l|}{\textbf{Relative FVC}} \\ \hline
 & <50 & 3.4\% \\ \hline
 & 50-69 & 37.0\% \\ \hline
 & 70-89 & 36.8\% \\ \hline
 & 90-109 & 15.2\% \\ \hline
 & 110+ & 7.6\% \\ \hline
\end{tabular}%
\end{table}

    A number of pre-processing steps were conducted to improve the consistency and quality of the CT images from the patient scans.  More specifically, all CT imaging data was translated to Hounsfield units (HU), and windowing was performed with a window level of -650 HU and a window width of 1700 HU to better focus on clinically relevant lung features.  Furthermore, synthetic padding and circular artifacts found in the CT imaging data for several patient cases within the patient cohort are mitigated to reduce the likelihood of erroneous visual features from being learned as predictive indicators.  Finally, calibration value errors found in the data for several patient cases within the patient cohort are accounted for to further mitigate the likelihood of erroneous characteristics from being learned as predictive indicators.  Example CT slices from the patient cohort are shown in Figure~\ref{fig:examples}. It can be observed that the visual appearance of pulmonary fibrosis in different patient CT scans can be quite varied, and thus can be quite challenging to utilize for lung function decline prediction.  The variable visual appearance in CT scans further motivate the exploration of deep learning strategies for tackling such a complex prediction task to facilitate for computer-assisted clinical decision support.

\subsection{Machine-driven design exploration}
    The goal of the proposed Fibrosis-Net is to predict the forced vital capacity (FVC) of a patient (in ml) at a specific time-point in the future given a patient's CT scan, initial spirometry measurement, and clinical metadata.  In order to construct a highly customized deep convolutional neural network architecture design tailored specifically for high predictive performance, we take inspiration from~\cite{covidnet} and a generative synthesis~\cite{gensynth} approach was leveraged as the machine-driven design exploration strategy for determining a strong architectural design for CT lung analysis.  In this approach, the problem of determining a tailored deep neural network architecture design is formulated as a constrained optimization problem based on a universal performance function $\mathcal{U}$ (e.g.,~\cite{wong2019netscore}) and a set of quantitative constraints. The aforementioned constrained optimization problem is solved in an iterative fashion, based on an initial network design prototype, the set of quantitative constraints, and data at hand, to produce network $N$.

    More specifically, the backbone architecture design for CT lung analysis identified via machine-driven design exploration leveraged residual architecture design principles~\cite{resnet,resnetv2} as an initial network design prototype, 2,116 patient cases acquired from around the world both with presence and absence of respiratory diseases for improve the quantity and diversity of CT scans, along with associated predictive performance constraints~\cite{covidnetct,covidnetct2}.  It is upon this backbone architecture design that the proposed Fibrosis-Net network architecture design was built to be tailored specifically for predicting FVC based on the CT scan, initial spirometry measurement, and clinical metadata of a patient.  The resulting Fibrosis-Net was implemented, trained, and evaluated using the TensorFlow deep learning library~\cite{tensorflow}, with training conducted using the Adam optimization algorithm~\cite{kingma2017adam} with a mean absolute error (MAE) loss, learning rate of 1e-4, exponential decay of 0.99 every 100 steps, and a batch size of 8.

        \begin{figure}[t]
        \centering
        \includegraphics[width=\textwidth]{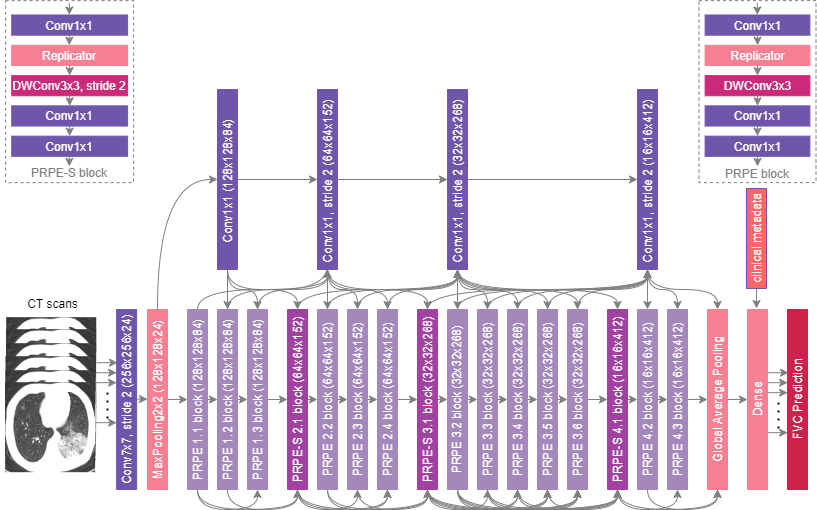}
        \caption{The proposed Fibrosis-Net architecture design. Given a patient's CT scan, initial spirometry measurement, and clinical metadata, the proposed Fibrosis-Net predicts the forced vital capacity (FVC) of a patient at a specified time-point in the future.  Fibrosis-Net exhibits an efficient network architecture design with light-weight components that strikes a strong balance between representational capacity and architecture and computational efficiency.}
        \label{fig:arch}
    \end{figure}

\subsection{Network architecture}
    The proposed Fibrosis-Net architecture is shown in Figure~\ref{fig:arch}.  Given a stack of CT images from a patient's CT scan, each CT image from the lower 55\% subset of the CT scan (where pulmonary fibrosis typically presents itself in the lungs) is passed through a series of convolutional layers to create a condensed feature representation characterizing the CT image.  This condensed feature representation of the CT image, along with clinical metadata, are then passed together into a dense layer to predict the linear rate of change in lung function.  The predicted linear rates of change in lung function from all of the CT images in a patient's CT scan are then passed into the FVC prediction layer, where the median predicted linear rate of change in lung function, the initial spirometry measurement of the patient, and clinical metadata (i.e., age, sex, and smoking status) are leveraged to predict the FVC at the desired time-point.  As the final operations in the FVC prediction layer, a regressor fitted on the clinical metadata with Elastic Net regularization is also leveraged to produce a predicted FVC at the desired time-point, which is then combined with the convolutional-driven FVC prediction to obtain the final FVC prediction.

    The proposed Fibrosis-Net possesses a highly efficient, heterogeneous design comprising largely of lightweight architectural components such as depthwise convolutions and pointwise convolutions.  In particular, similar to~\cite{covidnetct,covidnetct2}, both unstrided and strided projection-replication-projection-expansion design patterns (denoted as PRPE and PRPE-S for unstrided and strided patterns, respectively) are heavily utilized to strike a strong balance between representational capacity and architectural and computational efficiency.  The efficiency of the proposed Fibrosis-Net makes it very well-suited for clinical scenarios where computational resources are limited, particularly when dealing with CT scans consisting of many CT images such as the pulmonary fibrosis progression prediction task in this study.

    Furthermore, selective long-range connectivity is exhibited in the proposed Fibrosis-Net, with central connectivity hubs comprised of pointwise convolutions for enabling flow in information directly from earlier convolutional layers to the later convolutional layers.  By keeping the quantity of long-range connections very sparse through centralized connectivity hubs, the selective long-range connectivity characteristics of Fibrosis-Net strikes a strong balance between improved representational capacity and architectural efficiency.

    It can also be observed that Fibrosis-Net fuses learned visual features from a patient's CT scan with clinical metadata at a later stage of the network architecture design.  This enables the efficient utilization of important clinical knowledge captured within clinical metadata alongside important visual characteristics captured in a patient's CT scan for more well-informed, comprehensive predictive capabilities.  Finally, the utilization of a flexible FVC prediction layer in the proposed Fibrosis-Net architecture design takes into account a variable set of linear rates of change in lung function predicted by the dense layer depending on the quantity of CT images within a particular patient's CT scan.  As a result, this effectively allows for much greater flexibility in dealing with different real-world scenarios where the imaging protocol and imaging equipment parameters and configurations can vary greatly.

    These architectural characteristics possessed by the proposed Fibrosis-Net illustrates the effectiveness of leveraging machine-driven design exploration for constructing customized deep neural network architecture designs that are specifically tailored for clinical decision support tasks.

\subsection{Explainability-driven performance validation of Fibrosis-Net}\label{xai}
Understanding the behaviour of a deep neural network when making predictions is very important when used in clinical decision support scenarios, given that such predictions will impact patient care and influence treatment and management planning.  Inspired by this, we leverage an explainability-driven performance validation strategy to study the decision-making behaviour of Fibrosis-Net as to verify that predictions are based on relevant visual indicators in CT images.  Similar to~\cite{covidnet,covidnetct,covidnetct2,wong2021tbnet}, we leverage GSInquire~\cite{gsinquire} as the explainability method of choice for explainability-driven performance validation in this study to identify critical visual factors in CT images that Fibrosis-Net leverages to make FVC predictions.  More specifically, GSInquire harnesses the same generative synthesis strategy~\cite{gensynth} leveraged in the machine-driven design exploration process, as a previous study demonstrated the ability of GSInquire to provide explanations that better reflect the decision-making process of deep neural networks quantitatively when compared to other state-of-the-art explainability methods~\cite{gsinquire}.  More importantly, GSInquire is, to the best of the authors' knowledge, is one of the only explainability methods in literature that can be leveraged for studying and validating clinical regression problems, and thus makes it highly desirable for this study.

In brief, GSInquire utilizes the generative synthesis~\cite{gensynth} approach from the machine-driven design exploration process for identifying the backbone architecture design, where an inquisitor $\mathcal{I}$ is leveraged to probe a network $N$ with input $x$.  The reactionary responses $y$ from the probing process are leveraged by $\mathcal{I}$ to produce quantitative interpretation $z$ of the decision-making process of a deep neural network $N$ given an input $x$ in the same sub-space as $x$.  The details pertaining to GSInquire for explaining the decision-making behaviour of deep neural networks on CXR images can be found in~\cite{covidnet}.  An interesting property of GSInquire that also makes it well-suited for explainability-driven performance validation is that it is capable of producing explanations identifying specific critical factors within an image that quantitatively impacts the decisions made by a deep neural network, thus making it more readily interpretable and more quantitative for validation purposes than the types of relative importance variations visualized by other methods.

The details regarding how GSInquire can be leveraged to produce interpretations of deep neural network decision-making behaviour for clinical prediction tasks can be found in~\cite{covidnet}.  Here, the interpretation $z$ indicates the critical visual factors leveraged by Fibrosis-Net from CT images when making FVC predictions.

\section{Results and discussion}\label{results}

To explore and evaluate the efficacy of the proposed Fibrosis-Net for the prediction of lung decline progression due to pulmonary fibrosis from chest CT images, we take a multi-prone approach where we conduct: 1) an empirical quantitative performance evaluation of the deep neural network design to study its performance compared to last state-of-the-art methods, as well as 2) a visual validation evaluation of the decision-making behaviour of the deep neural network design using an explainability-driven performance validation process.  The quantitative and qualitative results are presented an discussed in detail below.

\subsection{Quantitative performance validation results}
    We quantitatively evaluate the efficacy of the proposed Fibrosis-Net using the OSIC Pulmonary Fibrosis Progression Challenge test cohort~\cite{OSIC}.  As consistent with the evaluation procedure described in~\cite{OSIC}, the modified Laplace Log Likelihood score is used, and a comparative analysis is conducted against the three Kaggle winning methods in the OSIC Pulmonary Fibrosis Progression Challenge.  The modified Laplace Log Likelihood score (denoted here as $L$) can be expressed as,
    \begin{equation}
    \sigma_{clipped} = max(\sigma, 70~ml),
    \end{equation}
    \begin{equation}
    \Delta = min ( |FVC_{true} - FVC_{predicted}|, 1000~ml),
    \end{equation}
        \begin{equation}
    L = -   \frac{\sqrt{2} \Delta}{\sigma_{clipped}} - \ln(\sqrt{2} \sigma_{clipped}).
     \end{equation}

\noindent The modified Laplace Log Likelihood score is negative in value and higher the score is, the better the performance of the method is for predicting pulmonary fibrosis progression.

    The modified Laplace Log Likelihood scores for the proposed Fibrosis-Net and the winning methods are shown in Table~\ref{table:comp}. It can be observed that Fibrosis-Net achieves a significantly higher modified Laplace Log Likelihood score when compared to the winning solutions in the challenge.  More specifically, Fibrosis-Net achieved a Laplace Log Likelihood score that exceeded the Kaggle 1st place winning solution by \textbf{0.0117}, which is significantly higher than the score gaps between the three winning solutions given the logarithmic scale of the modified Laplace Log Likelihood score.  Based on these results, it can be seen that Fibrosis-Net can achieve state-of-the-art performance for lung decline progression and demonstrates the efficacy of machine-driven design exploration for constructing deep neural network designs tailored for clinical decision support tasks.

    \begin{table}[t]
        \caption{Comparison of Laplace Log Likelihood scores for the Kaggle winning methods and the proposed Fibrosis-Net on the test cohort from the OSIC Pulmonary Fibrosis Progression Challenge. Best results highlighted in \textbf{bold}.}
        \medskip
        \label{table:comp}
        \centering
        \begin{tabular}{l|c}
            \toprule
            Method & Laplace Log Likelihood \\
            \midrule
            Kaggle 1st place~\cite{OSIC} & -6.8305 \\
            Kaggle 2nd place~\cite{OSIC} & -6.8311 \\
            Kaggle 3rd place~\cite{OSIC} & -6.8336 \\
                        \midrule
            Fibrosis-Net & \textbf{-6.8188} \\
            \bottomrule
        \end{tabular}
    \end{table}

\subsection{Explainability-driven performance validation results}
\begin{figure}[t]
        \centering
        \includegraphics[width=0.9\textwidth]{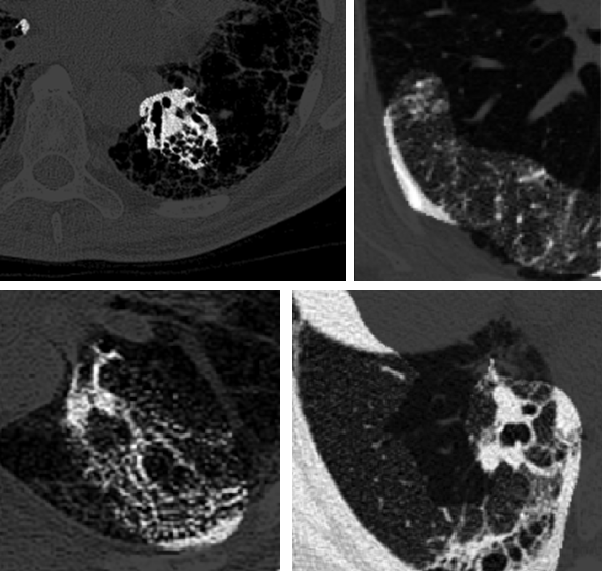}
        \caption{Example critical factors (highlighted as brighter regions) in example CT images of pulmonary fibrosis patients as identified by GSInquire~\cite{gsinquire}.  It can be observed that Fibrosis-Net is capable of leveraging clinically relevant visual indicators such as the presence and geographic extent of honeycombing in the lungs as presented in the CT images to make FVC predictions.}
        \label{fig:explain}
    \end{figure}

    As discussed earlier, we harnessed GSInquire~\cite{gsinquire} to conduct explainability-driven performance validation of Fibrosis-Net in order to study its behaviour when making predictions of lung function decline, as well as validate whether predictions are based on clinically-relevant imaging features rather than based on irrelevant features. Figure~\ref{fig:explain} illustrates example critical factors in CT images of pulmonary fibrosis patients as identified by GSInquire that are key to the decision-making behaviour of the proposed Fibrosis-Net.  It can be observed that Fibrosis-Net is capable of leveraging clinically relevant visual indicators such as the presence and geographic extent of honeycombing in the lungs as presented in the CT images to make FVC predictions. As such, it can be clearly seen that the proposed Fibrosis-Net is driven by correct, clinically relevant decision-making behaviour when making predictions of pulmonary fibrosis progression similar to those leveraged by clinicians~\cite{Devaraj}.  These visual results also highlight the importance of harnessing explainability-driven performance validation when building and evaluating deep neural networks for clinical decision support tasks.

    There are several important benefits to taking such an explainability-driven approach to performance validation, particularly for the proposed Fibrosis-Net where the purpose relates to clinical decision support.  First of all, by leveraging explainability-driven performance validation, one can obtain greater transparency and understanding into the decision-making behaviour of a deep neural network to ensure that it is leveraging clinically-relevant imaging features to make decisions (i.e., ``making the right decisions for the right reasons``).  Second, one can gain much greater insight into potential gaps, biases, and errors in both the data as well as in the decision-making behaviour of a deep neural network (i.e., ``making the right decisions for the wrong reasons`` based on irrelevant features such as synthetic padding, circular artifacts, etc.).  Third, by providing greater transparency into the decision-making processing during prediction, one can provide a greater sense of trust for clinicians leveraging such deep neural networks for computer-assisted clinical decision support and drive greater clinical adoption of such artificial intelligence-driven technologies.

Based on both quantitative and qualitative results, it was demonstrated that Fibrosis-Net can not only make FVC predictions at a higher level of accuracy than state-of-the-art methods, but also do it in a more trustworthy, validated manner that leverages clinically relevant visual indicators within the CT images of a pulmonary fibrosis patient.

\section{Conclusion and Discussion}
\label{conclusion}
In this study, we introduced Fibrosis-Net, a deep convolutional neural network design tailored for the prediction of pulmonary fibrosis progression from CT images. Designed with the help of a machine-driven design exploration strategy, Fibrosis-Net is available open source to the general public as part of the OpenMedAI initiative. Experimental results using a patient cohort from the OSIC Pulmonary Fibrosis Progression Challenge show that the proposed Fibrosis-Net can achieve state-of-the-art forced vital capacity prediction performance when compared to the winning solutions on the challenge. Furthermore, an explainability-driven performance validation of Fibrosis-Net showed that relevant visual indicators in the CT images were leveraged when producing predictions. Given the promise of the proposed Fibrosis-Net, we aim to explore this strategy for creating deep neural networks to perform other clinical decision support tasks for other pulmonary conditions such as chronic obstructive pulmonary disease prediction and pulmonary hypertension detection.

    Given the devastating effects of pulmonary fibrosis on a individual's health and well-being and the lack of a known cure, the research in lung function decline prediction presented in this study can have positive benefit to clinical scientists and researchers who are developing deep learning systems for supporting clinical workflows in a number of impactful ways. First, by illustrating the efficacy of machine-driven design for building highly tailored deep neural network architecture designs for a prediction task beyond the types of clinical decision support tasks illustrate in past studies~\cite{covidnet,covidnets,covidnetct,covidnetct2}, the hope is that other researchers and scientists may consider leveraging such an approach to accelerate and improve the design of deep learning solutions for different clinical scenarios.  Second, by illustrating the efficacy of explainability-based performance validation on gaining a better understanding of the behaviour of Fibrosis-Net on making FVC predictions, the hope is that other researchers and scientists may consider leveraging explainability methods more frequently to improve transparency and trust. Third and finally, the proposed Fibrosis-Net is released in an open-access, open-source fashion, thus allowing for researchers, scientists, and clinicians to leverage this work for further investigation and build upon this work to accelerate the development of clinically viable systems.

    It is important to note that Fibrosis-Net is by no means a production-ready clinical assessment solution, and is intended as a foundation for further research and development.  Furthermore, it is important to note that the predictions made by Fibrosis-Net and other similar artificial intelligence clinical decision support tools should not be accepted blindly but rather by utilized to aid clinicians in the clinical decision support process.  As such, much greater impact can be achieved with tools such as Fibrosis-Net are utilized in a human-in-the-loop manner.

\section*{Acknowledgement}
    We would like to thank Amazon Web Services, Open Source Imaging Consortium, and DarwinAI Corp.  We would also like to thank Hayden Gunraj from the University of Waterloo.

\section*{Author contributions statement}
A.W. conceived the experiments, J.L., A.D., P.M. M.F., D.M., J.H.L., and M.L. conducted the experiments, all authors analysed the results, and all authors reviewed the manuscript.

\section*{Declaration of interests}
A.W., J.L., A.D., P.M. M.F., D.M., J.H.L., and M.L. are affiliated with DarwinAI Corp.

\bibliographystyle{IEEEtran}
\bibliography{references}

\begin{thebibliography}{10}
\providecommand{\url}[1]{#1}
\csname url@samestyle\endcsname
\providecommand{\newblock}{\relax}
\providecommand{\bibinfo}[2]{#2}
\providecommand{\BIBentrySTDinterwordspacing}{\spaceskip=0pt\relax}
\providecommand{\BIBentryALTinterwordstretchfactor}{4}
\providecommand{\BIBentryALTinterwordspacing}{\spaceskip=\fontdimen2\font plus
\BIBentryALTinterwordstretchfactor\fontdimen3\font minus
  \fontdimen4\font\relax}
\providecommand{\BIBforeignlanguage}[2]{{%
\expandafter\ifx\csname l@#1\endcsname\relax
\typeout{** WARNING: IEEEtran.bst: No hyphenation pattern has been}%
\typeout{** loaded for the language `#1'. Using the pattern for}%
\typeout{** the default language instead.}%
\else
\language=\csname l@#1\endcsname
\fi
#2}}
\providecommand{\BIBdecl}{\relax}
\BIBdecl

\bibitem{Taniguchi}
H.~Taniguchi \emph{et~al.}, ``Pirfenidone in idiopathic pulmonary fibrosis,''
  \emph{European Respiratory Journal}, vol.~35, pp. 821--829, 2010.

\bibitem{Richeldi}
L.~Richeldi \emph{et~al.}, ``Efficacy and safety of nintedanib in idiopathic
  pulmonary fibrosis,'' \emph{The New England Journal of Medicine}, vol. 370,
  pp. 2071--2082, 2014.

\bibitem{Kistler}
K.~Kistler \emph{et~al.}, ``Lung transplantation in idiopathic pulmonary
  fibrosis: a systematic review of the literature,'' \emph{BMC Pulm Med.},
  vol.~14, p. 139, 2014.

\bibitem{raghu2018diagnosis}
G.~Raghu, M.~Remy-Jardin, J.~L. Myers, L.~Richeldi, C.~J. Ryerson, D.~J.
  Lederer, J.~Behr, V.~Cottin, S.~K. Danoff, F.~Morell \emph{et~al.},
  ``Diagnosis of idiopathic pulmonary fibrosis. an official ats/ers/jrs/alat
  clinical practice guideline,'' \emph{American journal of respiratory and
  critical care medicine}, vol. 198, no.~5, pp. e44--e68, 2018.

\bibitem{richeldi2017idiopathic}
L.~Richeldi, H.~R. Collard, and M.~G. Jones, ``Idiopathic pulmonary fibrosis,''
  \emph{The Lancet}, vol. 389, no. 10082, pp. 1941--1952, 2017.

\bibitem{tomassetti2016bronchoscopic}
S.~Tomassetti, A.~U. Wells, U.~Costabel, A.~Cavazza, T.~V. Colby, G.~Rossi,
  N.~Sverzellati, A.~Carloni, E.~Carretta, M.~Buccioli \emph{et~al.},
  ``Bronchoscopic lung cryobiopsy increases diagnostic confidence in the
  multidisciplinary diagnosis of idiopathic pulmonary fibrosis,''
  \emph{American journal of respiratory and critical care medicine}, vol. 193,
  no.~7, pp. 745--752, 2016.

\bibitem{Watters}
L.~Watters \emph{et~al.}, ``A clinical, radiographic, and physiologic scoring
  system for the longitudinal assessment of patients with idiopathic pulmonary
  fibrosis,'' \emph{American Review of Respiratory Disease}, vol. 133, no.~1,
  1985.

\bibitem{Russell}
A.~Russell \emph{et~al.}, ``Daily home spirometry: An effective tool for
  detecting progression in idiopathic pulmonary fibrosis,'' \emph{American
  Journal of Respiratory and Critical Care Medicine}, vol. 194, no.~8, 2016.

\bibitem{Wuyts}
W.~Wuyts \emph{et~al.}, ``Daily home spirometry: A new milestone in the field
  of pulmonary fibrosis,'' \emph{American Journal of Respiratory and Critical
  Care Medicine}, vol. 194, no.~8, 2016.

\bibitem{du2011forced}
R.~M. Du~Bois, D.~Weycker, C.~Albera, W.~Z. Bradford, U.~Costabel,
  A.~Kartashov, T.~E. King~Jr, L.~Lancaster, P.~W. Noble, S.~A. Sahn
  \emph{et~al.}, ``Forced vital capacity in patients with idiopathic pulmonary
  fibrosis: test properties and minimal clinically important difference,''
  \emph{American journal of respiratory and critical care medicine}, vol. 184,
  no.~12, pp. 1382--1389, 2011.

\bibitem{Devaraj}
A.~Devaraj, ``Imaging: how to recognise idiopathic pulmonary fibrosis,''
  \emph{European Respiratory Review}, vol.~23, pp. 215--219, 2014.

\bibitem{Gruden}
J.~F. Gruden, ``Ct in idiopathic pulmonary fibrosis: Diagnosis and beyond,''
  \emph{American Journal of Roentgenology}, vol. 206, pp. 495--507, 2016.

\bibitem{Souza}
C.~A. Souza, N.~L. Müller, J.~Flint, J.~L. Wright, and A.~Churg, ``Idiopathic
  pulmonary fibrosis: Spectrum of high-resolution ct findings,'' \emph{American
  Journal of Roentgenology}, vol. 185, pp. 1531--1539, 2005.

\bibitem{Lynch}
D.~Lynch, ``Ground glass attenuation on ct in patients with idiopathic
  pulmonary fibrosis,'' \emph{Chest Journal}, vol. 110, no.~2, pp. 312--313,
  1996.

\bibitem{covidnet}
L.~Wang, Z.~Q. Lin, and A.~Wong, ``Covid-net: A tailored deep convolutional
  neural network design for detection of covid-19 cases from chest x-ray
  images,'' \emph{Scientific Reports}, 2020.

\bibitem{covidnets}
A.~Wong, Z.~Q. Lin, L.~Wang, A.~G. Chung, B.~Shen, A.~Abbasi,
  M.~Hoshmand-Kochi, and T.~Q. Duong, ``Covid-net s: Towards computer-aided
  severity assessment via training and validation of deep neural networks for
  geographic extent and opacity extent scoring of chest x-rays for sars-cov-2
  lung disease severity,'' \emph{Scientific Reports}, 2021.

\bibitem{covidnetct}
H.~Gunraj and A.~Wong, ``Covidnet-ct: A tailored deep convolutional neural
  network design for detection of covid-19 cases from chest ct images,''
  \emph{Frontiers in Medicine}, 2020.

\bibitem{covidnetct2}
H.~Gunraj, A.~Sabri, D.~Koff, and A.~Wong, ``Covid-net ct-2: Enhanced deep
  neural networks for detection of covid-19 from chest ct images through
  bigger, more diverse learning,'' \emph{arXiv}, no. 2101.07433, 2021.

\bibitem{ebadi2021covidxus}
A.~Ebadi, P.~Xi, A.~MacLean, S.~Tremblay, S.~Kohli, and A.~Wong, ``Covidx-us --
  an open-access benchmark dataset of ultrasound imaging data for ai-driven
  covid-19 analytics,'' 2021.

\bibitem{lee2020cancernetsca}
J.~R.~H. Lee, M.~Pavlova, M.~Famouri, and A.~Wong, ``Cancernet-sca: Tailored
  deep neural network designs for detection of skin cancer from dermoscopy
  images,'' 2020.

\bibitem{wong2021tbnet}
A.~Wong, J.~R.~H. Lee, H.~Rahmat-Khah, A.~Sabri, and A.~Alaref, ``Tb-net: A
  tailored, self-attention deep convolutional neural network design for
  detection of tuberculosis cases from chest x-ray images,'' 2021.

\bibitem{OSIC}
\BIBentryALTinterwordspacing
OSIC, ``Osic pulmonary fibrosis progression,'' 2020. [Online]. Available:
  \url{https://www.kaggle.com/c/osic-pulmonary-fibrosis-progression/}
\BIBentrySTDinterwordspacing

\bibitem{Christe}
A.~Christe \emph{et~al.}, ``Computer-aided diagnosis of pulmonary fibrosis
  using deep learning and ct images,'' \emph{Investigative Radiology}, vol.~54,
  no.~10, pp. 627--632, 2019.

\bibitem{Walsh1}
S.~Walsh \emph{et~al.}, ``Deep learning for classifying fibrotic lung disease
  on high-resolution computed tomography: a case-cohort study,'' \emph{The
  Lancet Respiratory Medicine}, vol.~6, no.~11, 2018.

\bibitem{Walsh2}
------, ``A deep learning algorithm for classifying fibrotic lung disease on
  high resolution computed tomography,'' \emph{American Journal of Respiratory
  and Critical Care Medicine}, vol. 201, no. A7645, 2020.

\bibitem{Levin}
D.~Levin, ``Deep learning and the evaluation of pulmonary fibrosis,''
  \emph{Lancet Respir Med.}, vol.~6, no.~11, pp. 803--805, 2018.

\bibitem{anthimopoulos2016lung}
M.~Anthimopoulos, S.~Christodoulidis, L.~Ebner, A.~Christe, and S.~Mougiakakou,
  ``Lung pattern classification for interstitial lung diseases using a deep
  convolutional neural network,'' \emph{IEEE transactions on medical imaging},
  vol.~35, no.~5, pp. 1207--1216, 2016.

\bibitem{christodoulidis2016multisource}
S.~Christodoulidis, M.~Anthimopoulos, L.~Ebner, A.~Christe, and S.~Mougiakakou,
  ``Multisource transfer learning with convolutional neural networks for lung
  pattern analysis,'' \emph{IEEE journal of biomedical and health informatics},
  vol.~21, no.~1, pp. 76--84, 2016.

\bibitem{bermejo2020classification}
D.~Bermejo-Pel{\'a}ez, S.~Y. Ash, G.~R. Washko, R.~S.~J. Est{\'e}par, and M.~J.
  Ledesma-Carbayo, ``Classification of interstitial lung abnormality patterns
  with an ensemble of deep convolutional neural networks,'' \emph{Scientific
  reports}, vol.~10, no.~1, pp. 1--15, 2020.

\bibitem{tan2020efficientnet}
M.~Tan and Q.~V. Le, ``Efficientnet: Rethinking model scaling for convolutional
  neural networks,'' 2020.

\bibitem{Ekstrom}
E.~M, G.~T, B.~K \emph{et~al.}, ``Effects of smoking, gender and occupational
  exposure on the risk of severe pulmonary fibrosis: a population-based
  casecontrol study,'' \emph{BMJ Open}, vol.~4, 2014.

\bibitem{Kalafatis}
D.~Kalafatis, J.~Gao, I.~Pesonen \emph{et~al.}, ``Gender differences at
  presentation of idiopathic pulmonary fibrosis in sweden,'' \emph{BMC Pulm
  Med}, vol.~19, no.~22, 2019.

\bibitem{gensynth}
A.~Wong, M.~J. Shafiee, B.~Chwyl, and F.~Li, ``Ferminets: Learning generative
  machines to generate efficient neural networks via generative synthesis,''
  2018.

\bibitem{wong2019netscore}
\BIBentryALTinterwordspacing
A.~Wong, ``Netscore: Towards universal metrics for large-scale performance
  analysis of deep neural networks for practical usage,'' \emph{CoRR}, vol.
  abs/1806.05512, 2018. [Online]. Available:
  \url{http://arxiv.org/abs/1806.05512}
\BIBentrySTDinterwordspacing

\bibitem{resnet}
K.~He, X.~Zhang, S.~Ren, and J.~Sun, ``Deep residual learning for image
  recognition,'' in \emph{2016 IEEE Conference on Computer Vision and Pattern
  Recognition (CVPR)}, 2016, pp. 770--778.

\bibitem{resnetv2}
------, ``Identity mappings in deep residual networks,'' in \emph{Computer
  Vision - ECCV 2016}, B.~Leibe, J.~Matas, N.~Sebe, and M.~Welling, Eds.\hskip
  1em plus 0.5em minus 0.4em\relax Cham: Springer International Publishing,
  2016, pp. 630--645.

\bibitem{tensorflow}
M.~Abadi, A.~Agarwal, P.~Barham, E.~Brevdo, Z.~Chen, C.~Citro, G.~S. Corrado,
  A.~Davis, J.~Dean, M.~Devin, S.~Ghemawat, I.~Goodfellow, A.~Harp, G.~Irving,
  M.~Isard, Y.~Jia, R.~Jozefowicz, L.~Kaiser, M.~Kudlur, J.~Levenberg,
  D.~Man\'{e}, R.~Monga, S.~Moore, D.~Murray, C.~Olah, M.~Schuster, J.~Shlens,
  B.~Steiner, I.~Sutskever, K.~Talwar, P.~Tucker, V.~Vanhoucke, V.~Vasudevan,
  F.~Vi\'{e}gas, O.~Vinyals, P.~Warden, M.~Wattenberg, M.~Wicke, Y.~Yu, and
  X.~Zheng, ``{TensorFlow}: Large-scale machine learning on heterogeneous
  systems,'' 2015, software available from tensorflow.org.

\bibitem{kingma2017adam}
D.~P. Kingma and J.~Ba, ``Adam: A method for stochastic optimization,'' 2017.

\bibitem{gsinquire}
Z.~Q. Lin, M.~J. Shafiee, S.~Bochkarev, M.~S. Jules, X.~Y. Wang, and A.~Wong,
  ``Do explanations reflect decisions? a machine-centric strategy to quantify
  the performance of explainability algorithms,'' 2019.

\end{thebibliography}

\end{document}